\documentclass[12pt,preprint]{aastex}
\newcommand{\begit}{\begin{itemize}}
\newcommand{\enit}{\end{itemize}}
\newcommand{\begen}{\begin{enumerate}}
\newcommand{\enen}{\end{enumerate}}
\newcommand{\beq}{\begin{equation}} 
\newcommand{\eeq}{\end{equation}} 
\newcommand{\beqa}{\begin{eqnarray}} 
\newcommand{\eeqa}{\end{eqnarray}}

\begin{document} 

\title{Magnetic Protoneutron Star Winds and $r$-Process Nucleosynthesis}

\author{Todd A. Thompson\altaffilmark{1}}
\altaffiltext{1}{Hubble Fellow}
\affil{Astronomy Department and Theoretical Astrophysics Center, 
601 Campbell Hall, \\
The University of California, Berkeley, CA 94720;
thomp@astro.berkeley.edu}

\begin{abstract}
Because of their neutron-richness and association with supernovae,
post-explosion protoneutron star winds are thought to be a likely astrophysical
site for rapid neutron capture nucleosynthesis (the $r$-process).
However, the most recent models of spherical neutrino-driven 
protoneutron star winds do not produce robust $r$-process 
nucleosynthesis for `canonical' neutron stars with a gravitational 
mass of 1.4\,M$_\odot$ and coordinate radius of 10\,km.
These models fail variously; either the flow entropy is too low,
the electron fraction is too high, or the dynamical expansion timescale
is too long. To date, no models have included the effects of
an ordered dipole magnetic field.
We show that a strong magnetic field can trap the outflow
in the neutrino heating region, thus leading to much higher matter entropy.
We estimate both the trapping timescale and the resulting entropy amplification.
For sufficiently large energy deposition rates, the trapped matter 
emerges dynamically from the region of closed magnetic field lines and escapes to infinity.
We find that ordered dipoles with surface fields of 
$\gtrsim6\times10^{14}$\,G increase the asymptotic entropy
sufficiently for robust $r$-process nucleosynthesis.  
\end{abstract}


\section{Introduction}
\label{section:intro}

In the $r$-process, seed nuclei capture neutrons on timescales much shorter than those
for $\beta$-decay.  For sufficiently high neutron-to-seed ratio ($\gtrsim100$), 
nucleosynthesis proceeds to the heaviest nuclei, forming characteristic abundance 
peaks at $A\sim$80, 130, and 195 (Burbidge et al.~1957; Cameron 1957).
Observational studies of the relative abundance pattern of $r$-process elements in 
ultra-metal-poor ([Fe/H]$\lesssim-2.5$) halo stars reveal remarkable concordance 
with solar $r$-process abundances for $A\gtrsim135$ (e.g.~Sneden et al.~1996; 
Burris et al.~2000).  These observations suggest an astrophysical 
$r$-process site that consistently produces the same relative abundance in this mass range, 
at very early times in the chemical enrichment history of the galaxy. 
Although the governing nuclear physics is fairly well understood, 
the astrophysical site for the $r$-process has yet to be established.
Because of their intrinsic neutron-richness, the post-explosion outflows 
expected from protoneutron stars (PNSs) in the seconds 
after successful supernovae (SNe) are
thought to be a likely candidate site for the $r$-process 
(Woosley \& Hoffman 1992; Woosley et al.~1994; Burrows, Hayes, \& Fryxell 1995).  

The characteristics of any potential astrophysical $r$-process
environment that determine the resulting nucleosynthesis are the asymptotic entropy ($s_{\rm a}$;
units of $k_{\rm B}$\,baryon$^{-1}$ throughout), 
the electron fraction ($Y^{\rm a}_e$), and the dynamical timescale ($\tau_{\rm dyn}$, the 
density $e$-folding time at $T=0.5$\,MeV) near the start of the $r$-process.  
In general, higher $s_{\rm a}$, lower $Y^{\rm a}_e$, and shorter 
$\tau_{\rm dyn}$, lead to larger neutron-to-seed ratios and higher maximum $A$
(e.g.~Meyer \& Brown 1997).  In addition to attaining
the requisite $s_{\rm a}$, $Y^{\rm a}_e$, and $\tau_{\rm dyn}$, a potential 
$r$-process site must produce a sufficient quantity of $r$-process ejecta 
($M_{\rm ej}^{\rm r}$) per event, so that the total galactic $r$-process budget
is accounted for.  For example, if all supernovae produce a robust $r$-process
abundance pattern, then the total mass of $r$-process material that must be produced and 
ejected per SN is $M^{\rm r}_{\rm ej}\sim10^{-5}-10^{-6}$\,M$_\odot$ (Qian 2000).  

Recent PNS {\it wind} models (not {\it bubble} models, as 
in Woosley et al.~1994) fail to produce robust $r$-process nucleosynthesis up to and 
beyond the 3$^{\rm rd}$ $r$-process peak for `canonical' neutron stars with $M=1.4$\,M$_\odot$ 
and $R=10$\,km (Takahashi, Witti, \& Janka 1994; Qian \& Woosley 1996 (QW); Otsuki et al.~2000; 
Wanajo et al.~2001; Thompson, Burrows, \& Meyer 2001 (TBM); however, see Terasawa et al.~2002).  
Indeed, given the 
$Y^{\rm a}_e$ and $\tau_{\rm dyn}$ derived in these models, most efforts fall short of the 
required $s_{\rm a}$ by a factor of $3-5$.  

To date, protoneutron star wind models have neglected the effects of an ordered dipole
magnetic field.  Nagataki \& Kohri (2001) considered the effects of rotation and 
a monopole-like magnetic field on PNS winds in the spirit of Weber \& Davis
(1967), but were unable to explore field strengths above $\sim$10$^{11}$\,G 
because of the complicated critical point topologies encountered in that 
steady-state solution.  Although QW speculated on the role
of magnetic fields in their wind solutions, the effects discussed were not quantified.

Motivated by multi-dimensional studies of solar coronal hole and helmet streamers,
as well as by the very high inferred magnetic field strengths of some neutron stars
({\it magnetars}, Kouveliotou et al.~1999; Duncan \& Thompson 1992),
we here consider the effects of an ordered dipole magnetic field on
PNS winds and the composition of their ejecta. 
In this preliminary investigation, we show that strong dipole fields can trap
and heat matter in regions of closed magnetic field lines and that this matter
may be ejected to infinity with significantly amplified entropy.  Our estimates
indicate that for magnetar-strength initial dipole fields, the entropy increase
is sufficient for a robust $r$-process.

\section{Magnetic Protoneutron Star Winds}
\label{section:magnetic}

The comparison germane for assessing the importance of a dipole magnetic field 
on the wind evolution is the ratio $\beta=P/(B^2/8\pi)$, between the thermal pressure
and the magnetic energy density.
In Fig.~\ref{plot:pressr} we compare $P(r)$ for two 
wind models with $R=10$\,km, $M=1.4$\,M$_\odot$, and
$L_\nu^{\rm tot}=3.7\times10^{51}$\,erg\,s$^{-1}$ (labeled, `High
$L_\nu$') and $L_\nu^{\rm tot}=1.8\times10^{50}$\,erg\,s$^{-1}$ 
(`Low $L_\nu$') with $B^2/8\pi$ (dashed lines) for various surface field strengths ($B_0$).   
The spherical wind profiles shown here are obtained by solving the 
equations of steady-state general-relativistic hydrodynamics 
in a Schwarzschild metric with a general neutrino energy deposition function 
using a two-point relaxation algorithm on an adaptive radial mesh (as formulated in TBM).
In Fig.~\ref{plot:pressr}, we take $B(r)=B_0(R_B/r)^3$, where $R_B$ is a reference 
radius for the magnetic field footpoints.  Because of the exponential near-hydrostatic 
atmosphere, we set $R_B=11$\,km. Fig.~\ref{plot:pressr} shows that as the PNS cools
and $L_\nu$ decays, the dynamics of PNS winds can be dominated by magnetar-strength magnetic fields.
For $B_0<10^{14}$\,G, the magnetic field will only dominate the flow at very late times,
when both $L_\nu$ and the mass outflow rate ($\dot{M}$) are very small.
However, for $B_0=10^{15}$\,G, the magnetic field dominates even the High $L_\nu$
wind solution for $r<R_\beta\simeq30$\,km, where the relationship $\beta(R_\beta)=1$ defines $R_\beta$.
For small $B_0$, such that $R_\beta$ does not exist, the field is blown 
into a roughly monopolar configuration and the mass flow proceeds to infinity unencumbered.
However, for high $B_0$, we expect the global configuration to be much like 
that derived in multi-dimensional studies of solar coronal holes and helmet streamers 
(e.g.~Pneumann \& Kopp 1971; Steinolfson, Suess, \& Wu 1982; Usmanov et al.~2000; 
Lionello et al.~2002): open magnetic field lines at the poles and a region of closed 
magnetic field lines (a {\it dead zone}, in the spirit of Mestel 1968) 
at latitudes near the magnetic equator. 
 
In the closed zone the matter is trapped.
In the absence of net heating, this configuration might be long-lived.
In the context of PNS winds, however, we expect the pressure of the trapped matter to increase
as a result of neutrino energy deposition ($\dot{q}$).
If $P$ approaches $B^2/8\pi$ in the closed zone, we expect the matter previously trapped 
to escape dynamically.  Thus, given $\beta<1$ at $r\,(<R_\beta)$
and a local energy deposition rate $\dot{q}\rho$ [erg\,cm$^{-3}$\,s$^{-1}$],
the amount of time required to increase $P$ to $B^2/8\pi$ is 
\begin{equation}
\tau_{\rm trap}\sim[B^2/8\pi-P]/[\dot{q}\rho].
\label{tautrap}
\end{equation}
Hence, the matter in the closed zone is not trapped permanently, but
escapes in $\tau_{\rm trap}$ because of neutrino heating.  The
increase in $P$ of the trapped gas must be attended by an increase 
in entropy on account of $\dot{q}$. If $\dot{q}$ and $T$ do not change 
appreciably in $\tau_{\rm trap}$, $s_{\rm a}$ increases by 
\begin{equation}
\Delta s\sim\dot{q}\tau_{\rm trap}/T.
\label{entropyamp}
\end{equation}
If the magnetic field is very weak then $\tau_{\rm trap}$ is very short and $\Delta s\sim0$.  
At the other extreme, if $\beta\ll1$, some of the matter in the closed zone may be
permanently trapped (barring MHD instabilities, see \S\ref{section:questions}) because as $T$ gets large, 
neutrino cooling begins to compete with heating and $\dot{q}\rightarrow0$ locally.
The neutrino cooling rates are stiff functions of $T$ ($\propto T^6$ and $\propto T^9$
for $e^{-}p\rightarrow n\nu_e$ and $e^+e^-\rightarrow\nu\bar{\nu}$, respectively)
and for this reason, $T$, and, hence, $P$, cannot increase arbitrarily throughout the closed zone.
This sets a maximum $P$ increase attainable at any radius.
For example, Fig.~\ref{plot:temp} shows $T$, $s$, and $\rho$ (thin solid lines) for
the High $L_\nu$ model.  Also shown are the maximum temperature achievable at any radius
($T^\star$, dotted line) for fixed $\rho$ and the corresponding maximum entropy 
($s^\star$, dotted line).  The thick solid line shows $[s+\Delta s]$ as computed from 
eq.~(\ref{entropyamp}), for $B_0=5\times10^{15}$\,G.  
Because $[s+\Delta s]$ is greater than $s^{\star}$ for
$r<R_q$, the material inside $R_q$ is permanently trapped.  This defines $R_q$.
Note that in all models $R_q<R_\beta$.
Figure \ref{plot:temp} shows that although for very high $B_0$ some material may be 
permanently trapped, the material exterior to $R_q$ and interior to $R_\beta$ may escape
with significantly enhanced entropy.  In this case, near $R_q$, $s_{\rm a}$ is increased
from $\sim80$ to $\sim600$.
  
\section{Results}
\label{section:results}

Figure \ref{plot:ts} shows $s_{\rm a}$ versus $\tau_{\rm dyn}$
for a 10\,km 1.4\,M$_\odot$ PNS, for many different
$L_\nu$.  Each point on the thick solid line is a separate
spherical steady-state wind model with no magnetic field effects. 
As $L_\nu$ decreases, the curve evolves from low entropy 
($\sim$80) and short $\tau_{\rm dyn}$ ($\sim$4\,ms)
to higher entropy ($\sim$175) and considerably
longer $\tau_{\rm dyn}$ ($\sim$170\,ms).  One can think of this line 
as the cooling curve of a PNS composed of snapshots.  
The dashed line shows the analytical results from Hoffman, Woosley, \& Qian (1997) 
for the $s_{\rm a}$ required for 3$^{\rm rd}$-peak $r$-process nucleosynthesis
at fixed $Y_e^{\rm a}$ and $\tau_{\rm dyn}$.
Above this line, for $Y^{\rm a}_e=0.48$, 3$^{\rm rd}$-peak $r$-process elements can be synthesized.
Below it, the neutron-to-seed ratio is too small for the nuclear flow to proceed to $A\sim195$.  
One can see clearly that at all $\tau_{\rm dyn}$, $s_{\rm a}$
is too low to produce a robust $r$-process for the non-magnetic models.

The thin solid lines in Fig.~\ref{plot:ts} are obtained from the non-magnetic models
(thick solid line) by applying the procedure for calculating $\tau_{\rm trap}$ and
$\Delta s$ described in \S\ref{section:magnetic} at each $L_\nu$.  
Lines for surface magnetic field
strengths ($B_0$) of $2$, $4$, $6$, and $8\times10^{14}$\,G as well as 
$1$ and $2\times10^{15}$\,G are shown.  
We assume that $\tau_{\rm dyn}$ is the same between magnetic and non-magnetic models.
The increase in enthalpy due to trapping and consideration of the Bernoulli integral
indicate that this approximation should hold to no better than a factor of $2-3$.
For a given $B_0$, we find that if $R_\beta$ exists then $R_\beta\propto L_\nu^{-0.75}$ 
and $\tau_{\rm trap}\propto L_\nu^{-2.7}$,
to rough approximation.  One can see that for $B_0=2\times10^{14}$\,G 
$R_\beta$ does not exist until $\tau_{\rm dyn}\simeq0.075$\,s.
Figure \ref{plot:ts} shows that the entropy of the flow is profoundly effected by including
a strong dipole magnetic field. 
As long as $Y_e^{\rm a}<0.5$,  with a
surface field $\gtrsim6\times10^{14}$\,G, $s_{\rm a}$ may
be increased sufficiently for robust, 3$^{\rm rd}$-peak $r$-process nucleosynthesis.  

Because the solution to the wind equations constitutes an eigenvalue problem for
the mass outflow rate ($\dot{M}$), we can estimate the amount of 
$r$-process material ejected ($M_{\rm ej}^r$) by the magnetic models by 
assuming $L_\nu\propto t^{-\alpha}$. For a given $B_0$, $M_{\rm ej}^r$ is then the 
time integral of $\dot{M}$ above the dashed line in Fig.~\ref{plot:ts}.
Taking $\alpha=0.9$, the entire range of $L_\nu$ in Fig.~\ref{plot:ts}
corresponds to $\sim15$\,s.  Given this $L_\nu(t)$, 
$M_{\rm ej}^r(B_0=2\times10^{15}\,{\rm G})/
M_{\rm ej}^r(B_0=6\times10^{14}\,{\rm G})\sim50$.  
That is, for higher $B_0$, more $r$-process material is ejected.
Absolute numbers for $M_{\rm ej}^r$
are misleading because $\dot{M}\propto R^{5/3}$ 
(QW) and we have not included the PNS radial contraction 
during the early cooling epoch. However, as a first approximation, 
taking $R=30$\,km initially, 
with $R\propto t^{-1/3}$ (until $R$ reaches 10\,km),
with $L_\nu\propto t^{-0.9}$, and conserving magnetic flux so that when $R=10$\,km,
$B_0=5\times10^{15}$\,G, we find that $M_{\rm ej}^r\sim10^{-4}$\,M$_\odot$.
This number is sensitive to $R_q$, $R_\beta$, and the extent in 
latitude of the closed zone at the PNS surface, but we conclude that 
although the birth rate of neutron
stars with $B_0\sim10^{15}$\,G may be small compared to the total SN rate,
$M_{\rm ej}^r$ may be large enough to account for the lower rate
and accord with the total galactic $r$-process budget (see \S\ref{section:intro}).
That $M_{\rm ej}^r$ is not ejected 
isotropically may be important for the statistics of $r$-process enrichment in the early galaxy.

\section{Some Open Questions}
\label{section:questions}

{\bf Trapping Timescale:}
Equations (\ref{tautrap}) and (\ref{entropyamp}) are only applicable
if $\tau_{\rm trap}$ is much shorter than all other timescales in the
problem.  We expect global wind 
quantities to change on the timescale for neutrino diffusion and luminosity decay.
For example, the $B_0=2\times10^{15}$\,G line in Fig.~\ref{plot:ts} has
$\tau_{\rm trap}\simeq0.05$\,s and $\Delta s\simeq 90$ 
for the highest $L_\nu$ shown (shortest $\tau_{\rm dyn}$).  Because $\dot{q}$ and $\dot{M}$
do not change significantly over $\tau_{\rm trap}$, we expect the result presented in
Fig.~\ref{plot:ts} to be relatively robust.  However, taking
the $B_0=6\times10^{14}$\,G line in Fig.~\ref{plot:ts}, for the lowest $L_\nu$ 
(longest $\tau_{\rm dyn}$), we find that $\tau_{\rm trap}\simeq11$\,s and $\Delta s\simeq 650$.
In this case $\tau_{\rm trap}$ is nearly as long as the cooling/wind epoch itself and
the approximations of this paper break down.  

{\bf Complex Field Topologies:}
The magnetic field may be a complex of higher order multipoles
with a variety of local field strengths in this very early phase of PNS
evolution.  In this case we may consider a distribution of closed loops on 
the PNS surface.  Qualitatively, this should not alter the conclusions of this paper 
- there would simply be a distribution of $\tau_{\rm trap}$ and corresponding $\Delta s$ 
locally.  For $M_{\rm ej}^r$, we need only estimate the covering fraction
of closed loops.
Interestingly, such considerations lead one to consider magnetic reconnection
as a possible energy deposition mechanism (QW).  Naively, including such a term in
$\dot{q}$ may increase the entropy of the affected matter.  
QW and TBM
showed that $\sim10$\% increases to the total energy deposition rate 
can significantly increase $s_{\rm a}$, but that 
the radial position of an enhancement in $\dot{q}$ 
is paramount to determining whether or not the entropy is increased 
asymptotically in the flow.

{\bf MHD Instabilities:}
The estimate of $\Delta s$ in eq.~(\ref{entropyamp})
relies on the stability of the closed region over a time $\tau_{\rm trap}$.
Given the action of neutrino heating and the geometry of the closed zone,
it is possible that MHD instabilities might disrupt the system.
If the timescale for instabilities to grow 
is much shorter than $\tau_{\rm trap}$ {\it and} the instability allows
for the rapid removal of matter from the closed zone, the entropy
enhancements shown in Fig.~\ref{plot:ts} may be significantly decreased.  
Because of line-tying we expect global interchange instabilities to be suppressed
(J.~Arons, private communication).
Ballooning modes at $R_\beta$ may develop rapidly, but 
allow for only a small amount of matter to escape in $\tau_{\rm trap}$.
Perhaps such instabilities are the only way to remove the high entropy
matter interior to $R_q$ (see \S\ref{section:magnetic}).
A detailed stability analysis, or the solution to the full time-dependent
MHD problem is required.

{\bf Rotation \& Convection:}
Similar to MHD instabilities, rapid differential rotation or convection
may disrupt the global magnetic field configuration on timescales much 
shorter than $\tau_{\rm trap}$, and thereby compromise the entropy amplifications 
calculated here.
Even the shortest significant $\tau_{\rm trap}$ calculated in Fig.~\ref{plot:ts}
is $50$\,ms in duration, which may be much longer than the
initial PNS rotation and convective overturn timescales
(Duncan \& Thompson 1992; Thompson \& Duncan 1993).  

{\bf Spindown:} 
The wind itself may slow the PNS rotation,
carrying away angular momentum as the PNS contracts and cools.  
The PNS spins down on a timescale $\tau_J=J/\dot{J}\sim (3/5)MR^2/\dot{M}R_\beta^2$ 
(e.g.~Lamers \& Cassinelli 1999; Weber \& Davis 1967).  Taking the `Low $L_\nu$' 
solution ($\dot{M}\sim10^{-7}$\,M$_\odot$\,s$^{-1}$)  and 
$B_0=10^{16}$\,G, we find $R_\beta\sim2000$\,km and $\tau_J\sim200$\,s -- too
long in comparison with all other relevant timescales to influence the 
PNS period significantly.  For constant $R$, we see that for $L_\nu\propto t^{-\alpha}$
and $R_\beta\propto L_\nu^{-0.75}$ that $\tau_J\propto t^\alpha$; the increase in $R_\beta$
as $L_\nu$ decreases is insufficient to counter the rapidly decreasing mass flux. 
Of course, this neglects the coupling between $B_0$ and the rotation of the PNS, 
which likely exists if the magnetic field is generated by either a standard or 
turbulent/convective  dynamo (Thompson \& Duncan 1993; Thompson \& Murray 2001).
These estimates for $\tau_J$, however, assume the definition for $R_\beta$ in 
\S\ref{section:magnetic} and that $B\propto r^{\,-3}$.  Modifications to these 
assumptions change $\tau_J$ significantly.  

\section{Summary, Conclusions, \& Implications}

For the first time, we consider the effect of a strong dipole magnetic field
on protoneutron star winds and the resulting nucleosynthesis in their ejecta.
We show that magnetar-like field strengths can dominate the wind dynamics at 
all times during the PNS cooling epoch (see Fig.~\ref{plot:pressr}).
We argue that at radii close to the PNS, matter 
may be trapped by closed magnetic field lines where $\beta<1$.
The matter in any closed region is not trapped permanently, but escapes on
a timescale $\tau_{\rm trap}$, as a result of neutrino heating and the
necessarily commensurate increase in both pressure and entropy.  
As $\beta$ approaches unity in the closed zone, the material escapes dynamically. 
In Fig.~\ref{plot:ts} we 
show that for surface magnetic field strengths $\gtrsim6\times10^{14}$\,G,
the entropy enhancement engendered by this trapping effect is sufficient
to yield robust, 3$^{\rm rd}$-peak $r$-process nucleosynthesis.  In this
way, the factor of $\sim3-5$ in entropy previously unattained in spherical 
models of PNS winds may be achieved.  That the $r$-process
would occur at high entropy (and not, necessarily, at very short $\tau_{\rm dyn}$, as in TBM) 
in these models might help explain how, in the 
multi-dimensional parameter space of supernovae and PNS cooling,
the same $r$-process abundance profile above $A\sim135$ is observed in both
the sun and ultra-metal-poor halo stars (see \S\ref{section:intro}).
A close look at the models of Meyer \& Brown (1997) reveals that at $s_{\rm a}=500$,
for $\tau_{\rm dyn}\sim0.06$, there is considerable leeway in choosing $Y_e^{\rm a}$
(say 0.48 to 0.42) so that, approximately, the relative $r$-process abundances for 
$135\lesssim A \lesssim195$ accord well with those from the sun.  For a given $Y_e^{\rm a}$,
similar flexibility in $\tau_{\rm dyn}$ also exists at high $s_{\rm a}$
(also implied by results of Otsuki et al.~2002).

Clearly, much more work is required to understand wind
emergence, evolution, and nucleosynthesis during the cooling epoch of
highly magnetized protoneutron stars.  
This preliminary investigation is merely the first
step in quantifying one of the panoply of possible effects.
Although the magnitude of the entropy enhancement caused by trapping 
in closed magnetic loops is indicated in Fig.~\ref{plot:ts},
detailed numbers and systematics must await multi-dimensional MHD models.
That said, however, we here
forward the idea that neutrino-driven winds from nascent protoneutron stars with 
magnetar-like surface field strengths are responsible for the production of 
the heavy $r$-process nuclides we find in nature.

\acknowledgments

The author is indebted to Jon Arons, Eliot Quataert, and Adam Burrows
for a critical reading of the text and to Anatoly Spitkovsky, Stan Woosley, 
Lars Bildsten, and Henk Spruit for helpful conversations.
Insightful comments and suggestions for improvement of the manuscript 
by the referee, Al Cameron, are also acknowledged.
Support for this work was provided by NASA through Hubble Fellowship
grant \#HST-HF-01157.01-A awarded by the Space Telescope Science
Institute, which is operated by the Association of Universities for Research in Astronomy,
Inc., for NASA, under contract NAS 5-26555.

\figcaption{
Two representative wind pressure profiles 
(solid lines, $\log_{10}$[erg\,cm$^{-3}$])
as a function of radius for `High' and `Low' 
neutrino luminosity ($L_\nu$) for a neutron
star with gravitational mass 1.4\,M$_\odot$ 
and coordinate radius $R_\nu=10$\,km.  
Also shown are profiles of magnetic energy density 
($\log_{10}[B^2/8\pi\,\,({\rm erg}\,{\rm cm}^{-3})]$, dashed lines) 
assuming $B=B_0(R_B/r)^3$, where the reference 
radius $R_B=11$\,km, for surface magnetic field
strengths of 10$^{13}$, 10$^{14}$, 10$^{15}$, and 10$^{16}$\,G.
\label{plot:pressr}}

\figcaption{
$T$ (MeV), $s$, and $\rho$ ($\log_{10}[{\rm g\,\,cm}^{-3}]$),
for the `High $L_\nu$' wind solution (thin solid lines, compare with
Fig.~\ref{plot:pressr}).  For given $\rho(r)$, 
dotted lines show $T^*(r)$ and $s^*(r)$, the maximum 
temperature and corresponding entropy that can be attained before
cooling balances neutrino heating (see \S\ref{section:magnetic}).
For $B_0=5\times10^{15}$\,G, the thick solid line shows 
$s+\Delta s$, where $\Delta s$ is computed using eq.~(\ref{entropyamp}).
For this wind model and assumed $B_0$, 
$R_q\simeq23$\,km and $R_\beta\simeq125$.\label{plot:temp}}

\figcaption{
Asymptotic entropy ($s_{\rm a}$) versus 
dynamical timescale ($\tau_{\rm dyn}$) for a collection of 
PNS wind models with $R=10$\,km and $M=1.4$\,M$_\odot$ 
as a function of neutrino luminosity ($L_\nu$).  The thick solid line shows the 
evolution of the wind solutions in the $s_{\rm a}$-$\tau_{\rm dyn}$
plane without including any magnetic field effects.  Thin solid lines 
show the same curve, but including the amplification in $s_{\rm a}$ due 
to $B_0=2\times10^{14}$, $4\times10^{14}$, $6\times10^{14}$,
$8\times10^{14}$, $1\times10^{15}$, and $2\times10^{15}$\,G as 
per eq.~(\ref{entropyamp}). The heavy dashed line shows,
for $Y_e^{\rm a}=0.48$, the line in the 
$s_{\rm a}$-$\tau_{\rm dyn}$ plane above which
3$^{\rm rd}$-peak $r$-process nucleosynthesis 
is likely (from Hoffman, Woosley, \& Qian 1997).
\label{plot:ts}}

\newpage
\plotone{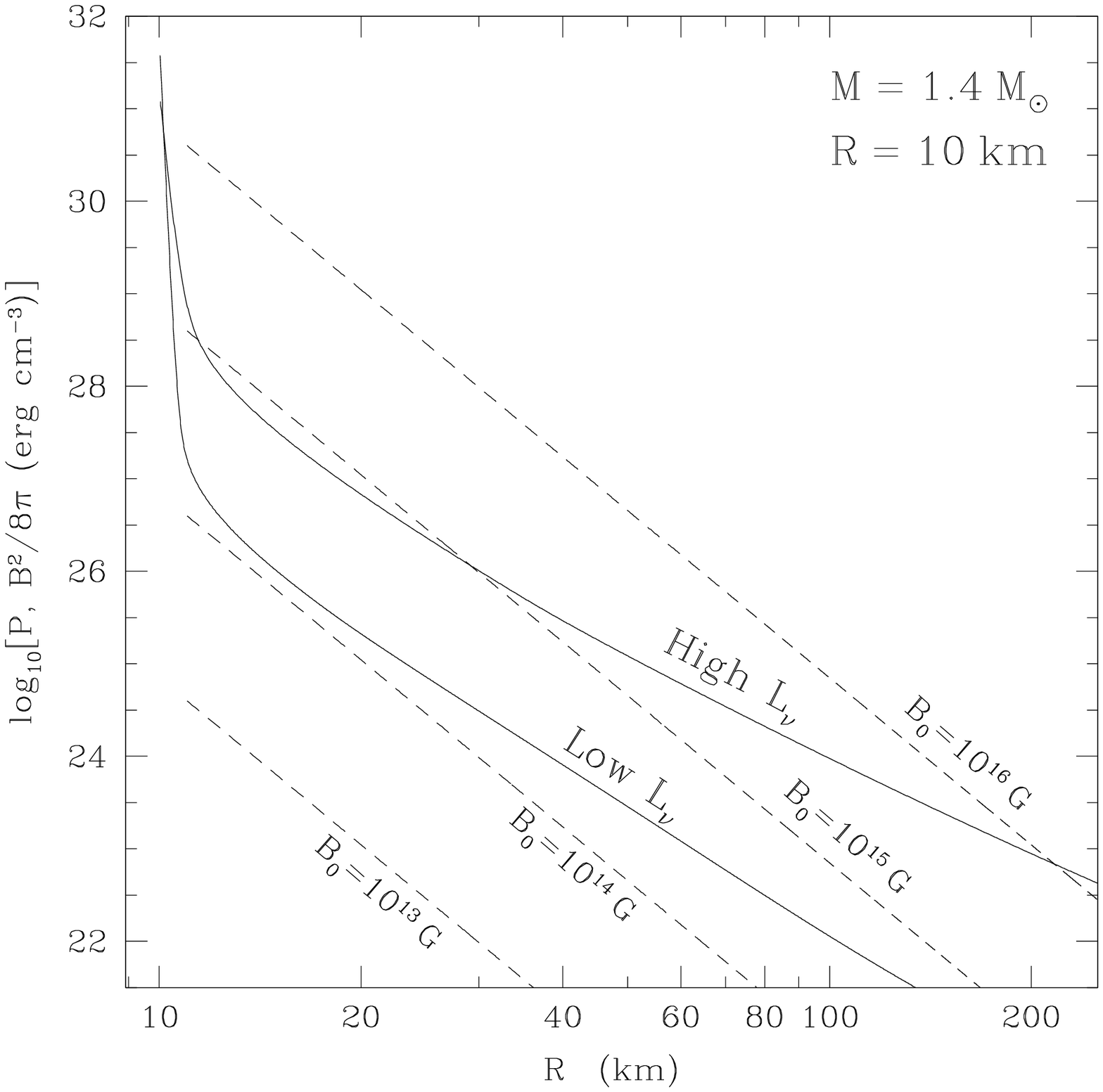}
\newpage
\plotone{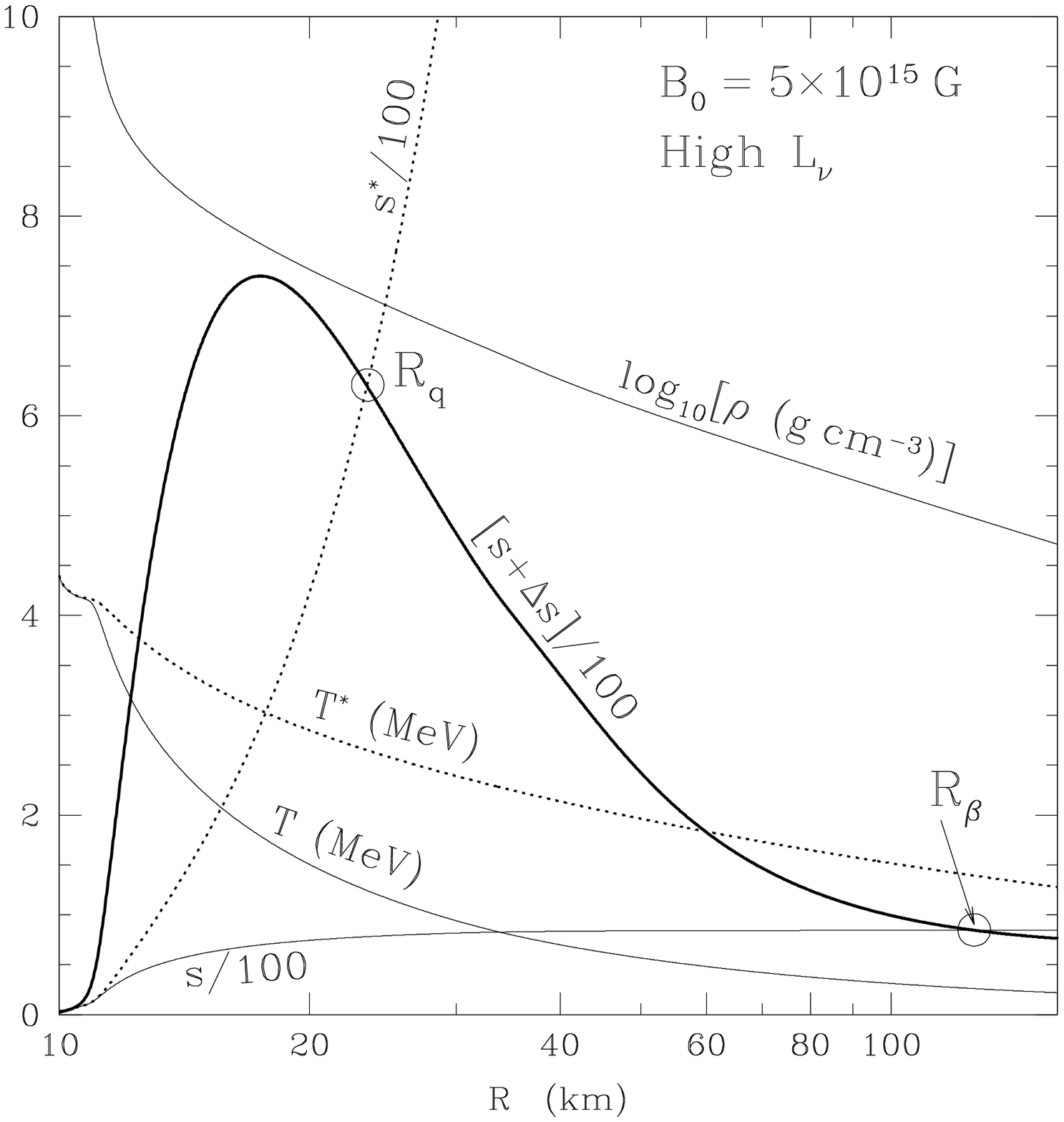}
\newpage
\plotone{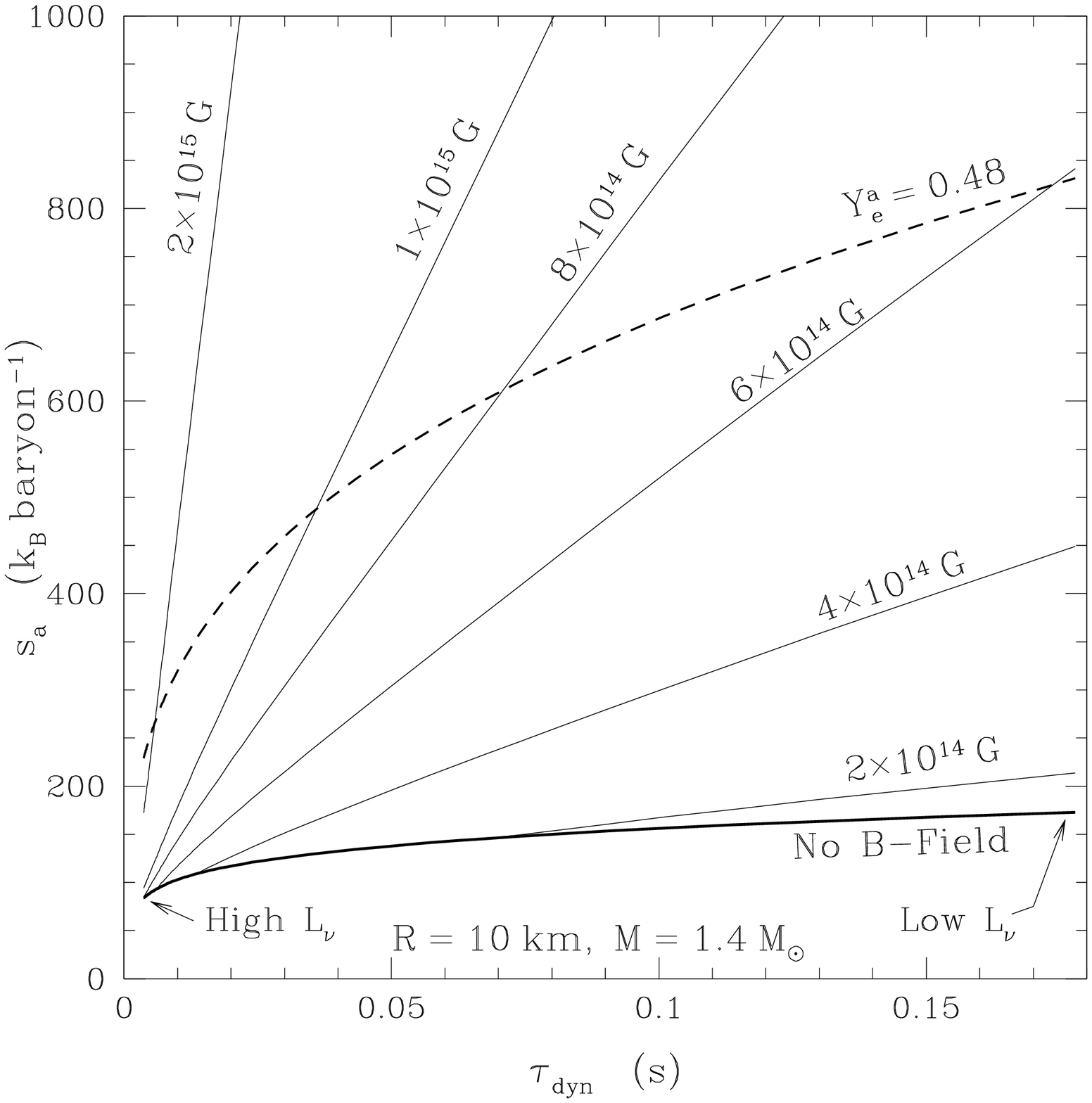}

\end{document}